\documentclass[10pt]{scrbook}

\usepackage{fontenc}

\usepackage[slantedGreek]{sfmath}

\usepackage[dvips]{graphicx}
\usepackage{color,amsmath,amssymb,amsfonts,mathrsfs,bbm}
\usepackage{afterpage,fancyhdr,tabularx}
\usepackage{geometry}
\usepackage{cite}

\usepackage{dpfloat}

\usepackage{color}
\definecolor{darkred}{rgb}{0.90,0,0}
\definecolor{darkgreen}{rgb}{0,0.60,.2}
\definecolor{darkblue}{rgb}{0,0,1}
\definecolor{grey}{cmyk}{0,0,0,0.25}
\definecolor{orange}{cmyk}{0,0.6,0.8,0}

\definecolor{blueP}{rgb}{0.0,0.08,0.45}

\usepackage[hang]{caption}

\DeclareCaptionFont{blueP}{\color{blueP}}
\usepackage{titlesec}
\titleformat{\chapter}[display]{\normalfont\huge\bfseries\color{blueP}}{\chaptertitlename\ \thechapter}{20pt}{\huge}
\titleformat{\section}{\normalfont\Large\bfseries\color{blueP}}{\thesection}{0.8em}{}
\titleformat{\subsection}{\normalfont\large\bfseries\color{blueP}}{\thesubsection}{0.8em}{}
\titleformat{\subsubsection}{\normalfont\normalsize\bfseries\color{blueP}}{\thesubsubsection}{1em}{}
\titlespacing*{\chapter}{0pt}{50pt}{20pt}
\titlespacing*{\section}{0pt}{2.75ex plus 0.7ex minus .2ex}{1.75ex plus .2ex}
\titlespacing*{\subsection}{0pt}{2.5ex plus 0.7ex minus .2ex}{1.2ex plus .2ex}
\titlespacing*{\subsubsection}{0pt}{2ex plus 0.7ex minus .2ex}{1ex plus .2ex}

\numberwithin{equation}{section}

\newcommand{\monthword}[1]{\ifcase#1\or Januar\or Februar\or M\"arz\or April\or
                                        Mai\or Juni\or Juli\or August\or
                                        September\or Oktober\or November\or Dezember\fi}

\makeatletter
\newcommand{\fmslash}[2][0mu]{%
 \mathchoice
   {\fmsl@sh\displaystyle{#1}{#2}}%
   {\fmsl@sh\textstyle{#1}{#2}}%
   {\fmsl@sh\scriptstyle{#1}{#2}}%
   {\fmsl@sh\scriptscriptstyle{#1}{#2}}}
\newcommand{\fmsl@sh}[3]{%
 \m@th\ooalign{$\hfil#1\mkern#2/\hfil$\crcr$#1#3$}}
\makeatother

\usepackage[  pdfdisplaydoctitle=true, pdfborder={0 0 0}, colorlinks=true, urlcolor=blueP,
              pdfpagelayout=SinglePage, pdfpagemode=UseOutlines, pdfstartview=Fit, 
pdftitle={The Functional Renormalization Group for Zero-Dimensional Quantum Systems in and out of Equilibrium},
pdfauthor={Christoph Karrasch},
pdfcreator={pdfTeX using libpoppler 3.1415926-1.40.9-2.2 (Web2C 7.5.7)}   ]{hyperref}

\begin{document}

\begin{titlepage}
\begin{center}
\sffamily {\LARGE \bfseries 
~\\
The Functional Renormalization Group\\[1ex] for Zero-Dimensional Quantum Systems\\[1.7ex] in and out of Equilibrium}\\[15ex]
\Large

Von der Fakult\"at f\"ur Mathematik, Informatik und Naturwissenschaften\\[0.5ex] der RWTH Aachen genehmigte Dissertation zur Erlangung \\[0.5ex] des akademischen Grades \textsl{Doctor rerum naturalium}\\[12ex]

von\\[2ex]
{\bfseries Dipl.-Phys.\ Christoph Karrasch}\\[2ex]
aus Duderstadt
{\\[16ex] \large 6.~September 2010 }\\[16ex]

\includegraphics[width=0.4\linewidth,clip]{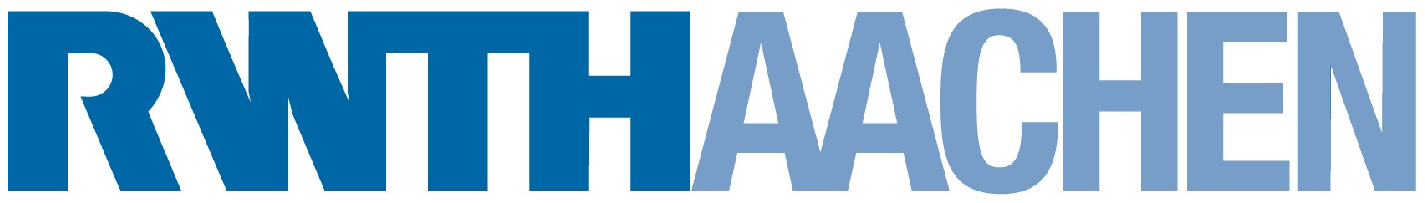}
\end{center}

%
%
%

\newpage
\thispagestyle{empty}\setlength{\parindent}{0cm}{\bfseries
Contact Information:\\\\\\
Christoph Karrasch\\\\
Institut f\"ur Theoretische Physik A\\
Physikzentrum, RWTH Aachen\\
52056 Aachen\\\\
phone: +49 241 8027032\\
email: karrasch (at) physik.rwth-aachen.de\\
homepage: http://www.theorie.physik.uni-goettingen.de/$\sim$karrasch\\[40ex]

\begin{minipage}[t]{0.39\linewidth}
Berichter:
\end{minipage}\begin{minipage}[t]{0.5\linewidth}
Volker Meden\\[1ex] Herbert Schoeller
\end{minipage}\\[4ex]
\begin{minipage}[t]{0.39\linewidth}
Tag der m\"undlichen Pr\"ufung:\end{minipage}\begin{minipage}[t]{0.5\linewidth}
2.~Juli 2010
\end{minipage}

}
\setlength{\parindent}{2ex}
\end{titlepage}

\pagestyle{empty}

\chapter*{Short Summary}
\subsubsection{Functional RG for Quantum Impurity Systems: Status Report}
In this Thesis, we study transport properties of quantum impurity systems using the functional renormalization group (FRG). The latter is an RG-based diagrammatic tool to treat Coulomb interactions in a more flexible (but less accurate) way than, e.g., by virtue of the numerical renormalization group approach. It was first applied to quantum dot systems, where electronic correlations lead to interesting strong-coupling effects, roughly five years ago. The employed approximation scheme, which can be viewed as a kind of RG enhanced Hartree Fock theory not suffering from typical mean-field artifacts, succeeds in accurately describing \textsl{linear transport properties} (such as the conductance) of various single- as well as multi-level spinful and spin-polarised quantum dot geometries at \textsl{zero temperature} and even captures aspects of Kondo physics [1-5].

\subsubsection{Functional RG for Quantum Impurity Systems: Goals}
In a nutshell, advance in this Thesis is three-fold. First, we introduce a \textsl{frequency-dependent second-order truncation scheme} in order to eventually address \textsl{finite-energy linear-response transport} properties of quantum dot systems. Secondly, a generalisation of the Hartree-Fock-like FRG approximation to Keldysh space allows for computing \textsl{non-linear steady-state transport} properties. Thirdly, we investigate the physics of a quantum dot Josephson junction as well as the charging of a single narrow level, (mainly) using the frequency-independent approach.

\subsubsection{Method Development, Vol.~I: Finite-Frequency Properties}
As mentioned above, the FRG was mainly used to compute equilibrium zero-energy properties of quantum dot systems (such as the linear conductance) in the $T=0$ -- limit. In order to treat finite temperatures and to address energy-dependent observables (such as the density of states), one needs to account for an additional higher-order class of functional RG flow equations -- which is technically involved. We demonstrate for two distinct problems (namely the single impurity Anderson as well as the interacting resonant level model) that this turns out to be possible in principle and leads to systematic improvements for small to intermediate Coulomb interactions [6,10]. In general, however, calculating energy-dependent properties needs for an ill-controlled analytic continuation of numerical Matsubara data which can only be circumvented in certain special situations [12]. More severely, aspects of Kondo physics contained in the simple Hartree-Fock-like functional RG approximation scheme can no longer be described by the -- \textsl{a priori} more elaborate -- higher-order approach. Thus, it is still an altogether open issue how to reliably compute energy-dependent properties (e.g., the density of states) in the strong-coupling limit using the functional RG.

\subsubsection{Method Development, Vol.~II: Towards Non-Equilibrium}
Treating systems in non-equilibrium requires a fundamental extension of the method to Keldysh space. This can be done straightforward in the long-time (steady-state) limit, and even the most simple (Hartree-Fock-like) FRG approximation scheme shows satisfying agreement with time-dependent density matrix renormalization group (DMRG) data published for the interacting resonant level model [10]. The latter provides a reasonable basis for a study of non-equilibrium transport through a quantum dot dominated by charge fluctuations. In the so-called scaling limit of large bandwidths (which cannot be addressed, e.g., by the DMRG), it features universal power laws which can be described analytically by the functional renormalization group scheme in complete agreement with real-time RG data [11].

\subsubsection{The Quantum Dot Josephson Junction}
The Josephson current through a quantum dot coupled to superconductors is governed by a singlet-doublet quantum phase transition. Experimental progress in realising such systems has triggered a lot of interest in modelling quantum impurities attached to BCS leads. In this line, the functional RG allows for calculating both the phase boundary and the supercurrent in good agreement with exact results as well as with numerical RG reference data [5]. Whereas the latter is accurate for arbitrarily large values of $U$ but limited to highly symmetric problems, any system parameters -- particularly the experimentally most relevant case of finite gate voltages [9] -- can be treated by the FRG approach. Placing the quantum dot in an interferometric Aharonov-Bohm geometry leads to multiple singlet-doublet transitions, and the model exhibits re-entrance behaviour [7].

\subsubsection{Charging of Narrow Quantum Dot Levels}
A quantum dot which comprises of one level (labelled by $\sigma=+$) contacted to a higher-dimensional bath by tunnel barriers of height $\Gamma_+$ as well as a second level ($\sigma=-$) that couples to the system via a Coulomb repulsion features a `quantum phase transition' as the energy of the latter crosses the chemical potential. In presence of small tunnelling elements to some bath ($\Gamma_-$) or to the first level ($t'$) -- which might be an overall generic scenario within various experimental situations -- the charging transition acquires a finite width scaling as a power law of the bare coupling strength $t',\Gamma_-\ll\Gamma_+$. This can be shown analytically by mapping the system to the anisotropic Kondo model using bosonisation (and exploiting well-known results for the latter). We confirm power-law variations using the functional and numerical renormalization group frameworks [5].

\vspace*{3ex}
\begin{enumerate}\setlength{\itemsep}{0.5em}
\item[{[1]}]
C.~Karrasch, T.~Enss, and V.~Meden\\[0.5ex]
\textsl{`A functional renormalization group approach to transport through correlated quantum dots'}\\[0.5ex]
\href{http://prb.aps.org/abstract/PRB/v73/i23/e235337}{Phys.~Rev.~B {\bfseries 73}, 235337 (2006)}
\item[{[2]}]
C.~Karrasch, T.~Hecht, A.~Weichselbaum, Y.~Oreg, J.~von Delft, and V.~Meden\\[0.5ex]
\textsl{`Mesoscopic to Universal Crossover of the Transmission Phase of Multilevel Quantum Dots'}\\[0.5ex]
\href{http://prl.aps.org/abstract/PRL/v98/i18/e186802}{Phys.~Rev.~Lett.~{\bfseries 198}, 186802 (2007)}
\item[{[3]}]
C.~Karrasch, T.~Hecht, A.~Weichselbaum, Y.~Oreg, J.~von Delft, and V.~Meden\\[0.5ex] 
\textsl{`Phase lapses in transmission through interacting two-level quantum dots'}\\[0.5ex]
\href{http://dx.doi.org/10.1088/1367-2630/9/5/123}{New.~J.~Phys.~{\bfseries 9}, 123 (2007)}
\item[{[4]}]
S.~Andergassen, T.~Enss, C.~Karrasch, and V.~Meden\\[0.5ex]
\textsl{`A gentle introduction to the functional renormalization group: The Kondo effect in quantum dots'}\\[0.5ex]
in ``Quantum Magnetism'' (Springer, New York, 2008)
\item[{[5]}]
C.~Karrasch, A.~Oguri, and V.~Meden\\[0.5ex]
\textsl{`Josephson current through a single Anderson impurity coupled to BCS leads'}\\[0.5ex]
\href{http://prb.aps.org/abstract/PRB/v77/i2/e024517}{Phys.~Rev.~B {\bfseries 77}, 024517 (2008)}
\item[{[6]}]
C.~Karrasch, R.~Hedden, R.~Peters, Th.~Pruschke, K.~Sch\"onhammer, and V.~Meden\\[0.5ex]
\textsl{`A finite-frequency functional renormalization group approach to the single impurity Anderson model'}\\[0.5ex]
\href{http://dx.doi.org/10.1088/0953-8984/20/34/345205}{J.~Phys.: Condensed Matter {\bfseries 20}, 345205 (2008)}
\item[{[7]}]
C.~Karrasch and V.~Meden\\[0.5ex]
\textsl{`Supercurrent and multiple singlet-doublet phase transitions of a quantum dot Josephson junction inside an Aharonov-Bohm ring'}\\[0.5ex]
\href{http://link.aps.org/abstract/PRB/v79/e045110}{Phys.~Rev.~B {\bfseries 79}, 045110 (2009)}
\item[{[8]}]
V.~Kashcheyevs, C.~Karrasch, T.~Hecht, A.~Weichselbaum, V.~Meden, and A.~Schiller\\[0.5ex]
\textsl{`Quantum Criticality Perspective on the Charging of Narrow Quantum-Dot Levels'}\\[0.5ex]
\href{http://prl.aps.org/abstract/PRL/v102/i13/e136805}{Phys.~Rev.~Lett.~{\bfseries 102}, 136805 (2009)}
\item[{[9]}]
A.~Eichler, R.~Deblock, M.~Weiss, C.~Karrasch, V.~Meden, C.~Sch\"onenberger, and H.~Bouchiat\\[0.5ex]
\textsl{`Tuning the Josephson current in carbon nanotubes with the Kondo effect'}\\[0.5ex]
\href{http://prb.aps.org/abstract/PRB/v79/i16/e161407}{Phys.~Rev.~B {\bfseries 79}, 161407(R) (2009)}
\item[{[10]}]
C.~Karrasch, M.~Pletyukhov, L.~Borda, and V.~Meden\\[0.5ex]
\textsl{`Functional renormalization group study of the interacting resonant level model in and out of equilibrium'}\\[0.5ex]
\href{http://prb.aps.org/abstract/PRB/v81/i12/e125122}{Phys.~Rev.~B {\bfseries 81}, 125122 (2010)}
\item[{[11]}]
C.~Karrasch, S.~Andergassen, M.~Pletyukhov, D.~Schuricht, L.~Borda, V.~Meden, and H.~Schoeller\\[0.5ex]
\textsl{`Non-equilibrium current and relaxation dynamics of a charge-fluctuating quantum dot'}\\[0.5ex]
\href{http://dx.doi.org/10.1209/0295-5075/90/30003}{Eur.~Phys.~Lett.~{\bfseries 90}, 30003 (2010)}
\item[{[12]}]
C.~Karrasch, V.~Meden, and K.~Sch\"onhammer\\[0.5ex]
\textsl{`Finite-temperature linear conductance from the Matsubara Green function without analytic continuation to the real axis'}\\[0.5ex]
\href{http://link.aps.org/doi/10.1103/PhysRevB.82.125114}{Phys.~Rev.~B {\bfseries 82}, 125114 (2010)}
\end{enumerate}

\vspace*{0.5cm}
\Large
For the complete Thesis, please go to

\vspace*{0.3cm}
\large
\href{http://www.theorie.physik.uni-goettingen.de/~karrasch/publications/thesis_karrasch.pdf}{www.theorie.physik.uni-goettingen.de/$\sim$karrasch/publications/thesis\_karrasch.pdf}

\vspace*{0.3cm}
\normalsize
(sorry for the inconvenience)

\end{document}